\documentclass{Interspeech}
\usepackage{multirow}
\usepackage{xcolor}
\usepackage{textcomp}
\usepackage{booktabs}
\usepackage{color, soul}
\usepackage{url,hyperref}
\interspeechcameraready

\title{Wanna hear your voice? A sample is all we need!
}

\author[affiliation={1}]{The Hieu}{Pham}
\author[affiliation={1}]{Phuong Thanh}{Tran Nguyen}
\author[affiliation={1}]{Xuan Tho}{Nguyen}
\author[affiliation={2}]{Tan Dat}{Nguyen}
\author[affiliation={1}]{Duc Dung}{Nguyen}

\affiliation{Ho Chi Minh city University of Technology}{VNUHCM
}{Vietnam}
\affiliation{Department of Electrical Engineering}{Korea Advanced Institute of Science and Technology (KAIST)}{Korea}
\email{\{hieu.pham14022003, thanh.tran159, tho.nguyenxuantho573\}@hcmut.edu.vn,
tandat.kaist@kaist.ac.kr,nddung@hcmut.edu.vn}

\usepackage{comment}

\begin{document}

\maketitle

\begin{abstract}
    
    Research on audio clue-based target speaker extraction (TSE) has focused on modeling mixtures and reference speech, achieving strong results in English due to abundant datasets. However, cross-lingual properties remain underexplored, as low-resource languages face challenges from limited annotated data and linguistic resources. To bridge this gap, we propose WHYV (Wanna Hear Your Voice), a cross-lingual TSE framework enabling zero-shot adaptation without fine-tuning. WHYV employs a frequency-modulated gating mechanism that dynamically adjusts the acoustic features of the target speaker, minimizing reliance on language-specific cues. Evaluations demonstrate state-of-the-art zero-shot performance: 13.8 dB (Libri2Mix mix-both), 18.1 dB (mix-clean), and 14.8 dB on Vietnamese data.
    We provide the live demo at: \url{https://anonymous.4open.science/w/WHYV/}
    
\end{abstract}

\section{Introduction}
\label{sec:intro}

Humans innately possess the ability to segregate various audio signals, particularly distinguishing between different speakers’ voices and separating speech from complex background noise.
Replicating this ability remains a significant challenge in the development of modern intelligent speech systems.
This widely recognized task is often referred to as the ``cocktail party problem''~\cite{Adelbert15cocktail_party,cherry1953some,haykin2005cocktail}.
Solving this issue would enable speech systems to be more robust, allowing them to effectively perform tasks such as Speech Recognition~\cite{raj2021integration,chen2021continuous}, Audio Source Separation~\cite{liu2022separate}, Data Collection~\cite{emilia}, and beyond. 

Current methodologies for addressing this challenge predominantly encompass three principal approaches: blind source separation (BSS), target speaker extraction (TSE), and noise reduction techniques.
Noise reduction is primarily applicable in scenarios where a single speech signal is foregrounded and remaining speech signals are relegated to background noise~\cite{Zmolikova23Neural}.
However, real-world scenarios of the ``cocktail party problem'' involve two or more overlapping speech sources within a single utterance, with close amplitude levels. Depending on the task requirements, speech systems must either separate multiple speakers’ voices simultaneously (BSS problem~\cite{Madhab2013BSS}) 
or extract only the target speaker’s speech (TSE problem~\cite{Zmolikova23Neural}). 

Recent research on BSS has achieved remarkable performance in speech separation tasks ~\cite{ZhongQiuWang2023TF_gridnet,shin2024SeparateandReconstruct,Subakan21Attention}, reaching over 20 dB in SI-SDRi on several benchmarks.
Moreover, some studies have attempted to address BSS with an unknown number of speakers~\cite{Lee2024boosting, Maiti2023EEND_SS}.
However, in many practical scenarios, such as personal speech systems, it is often necessary to extract the speech of a known target speaker while mitigating background noise and suppressing unexpected interfering voices~\cite{king2017robust,mallidi2018device,wang2018deep}.
In contrast to conventional BSS approaches, TSE mitigates several critical limitations, including mixing up speaker labels, relying on a fixed number of speakers, and tracking speakers over time. Instead, TSE focuses on accurately isolating the desired speaker's voice.
Recent TSE systems utilize various types of information to identify the target speaker, including location and direction \cite{Ge22L_spex}, visual cues \cite{Pan23ImagineNet}, and text input \cite{XiangHao2023typing,li24r_interspeech}, especially reference audio as an indicator of target speaker~\cite{ThaiBinh2024ConvoiFilter,kamo23_interspeech,Tuochao24Target,hu2024smma_net,Hiroshi24speakerbeam_ss,Veluri24Look_Once,Fengyuan24X_TF_GridNet}.

\begin{figure}[t]
  \centering
  \vspace{2mm}
  \includegraphics[width=.86\linewidth]{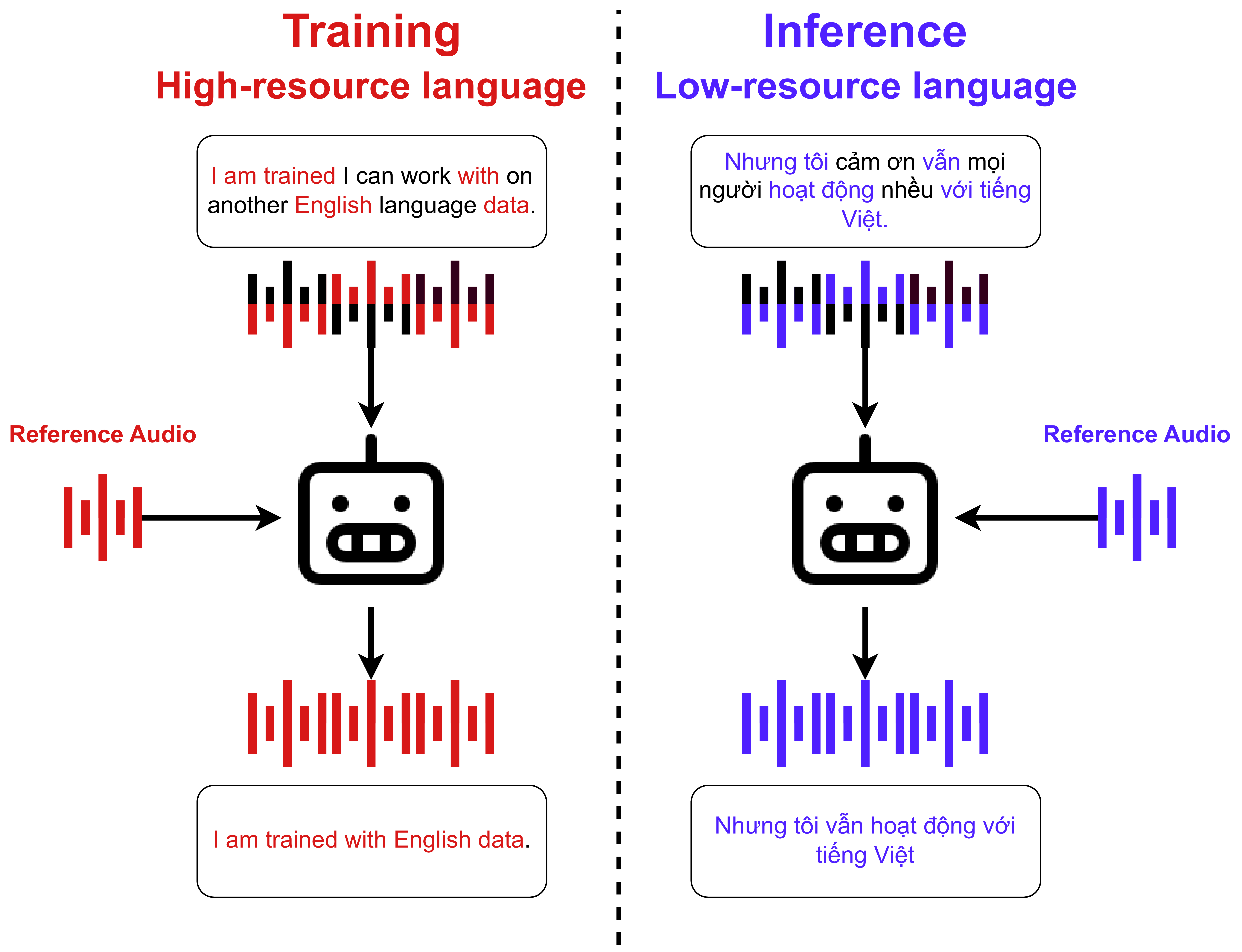}
  \caption{Visualize a zero-shot multi-language adaptive TSE}
  \label{fig:mullang}
  \vspace{-5 mm}
\end{figure}

However, most of these methods are 
predominantly optimized for English, with limited investigation into multilingual generalization of TSE.
Recent study~\cite{Tuochao24Target} on multilingual was conducted in English and Mandarin with limited performance. 
Furthermore, adapting these models to other languages poses significant challenges and high costs, particularly when the data source of the target language is insufficient. 
In this paper, we tackle this problem by introducing a straightforward yet efficient model called WHYV (Wanna Hear Your Voice). Building on the hypothesis that the acoustic properties of a voice are largely language-independent, we propose a novel Speaker Fusion Module that accurately isolates the target voice from mixed audio with high precision. 
Together with the proposed Global Target Filter (GTF) and Global Target Bias (GTB), the Speaker Fusion Module has successfully separated the voice of the reference speaker from the input mixture. 
Through extensive experiments, our model not only outperforms recent TSE models in the trained language evaluation, but also surpasses them in zero-shot adaptation to Vietnamese and Mandarin. Notably, this adaptation does not require an additional training stage. In addition, we also contribute a 13.4 hours evaluation dataset on Vietnamese to the community for language adaptation benchmark.

\section{WHYV Design}
\label{sec:method}
\subsection{Problem setting}
\label{sec:backgrounds}
Consider a single-channel time-domain mixture signal \( \mathbf{x} \in \mathbb{R}^L \) of length $L$, which contains speech from $C$ speakers along with ambient noise in an anechoic environment. The target speaker's speech, denoted as \( \mathbf{\hat{x}} \in \mathbb{R}^L \), can be expressed as follows:
\begin{equation}
\label{eq1}
\mathbf{x} = \mathbf{\hat{x}} + \sum_{k \in (0,C) \setminus \{s\}}{\mathbf{w_k}} + \mathbf{n},
\end{equation}
where \( \mathbf{w_k} \in \mathbb{R}^L\) represents the speech of the \( k \)-th speaker, \( k \) ranges over all speakers except the target speaker \( s \), and \( \mathbf{n} \in \mathbb{R}^L\) denotes the ambient noise. The target speaker extraction model $f$, parameterized by \(\theta\) and incorporating target speaker's information $c$, is defined as:
\begin{equation}
    \mathbf{\hat{x}} = f(\mathbf{x}, c; \theta).
\end{equation}
\subsection{Model architecture}
\subsubsection{Separation Module}
TF-GridNet \cite{ZhongQiuWang2023TF_gridnet} is used as the main backbone, serving as the primary separation module in the WHYV network. By leveraging a time-frequency domain representation, TF-GridNet captures both spectral and temporal features of the signal \cite{MEHRISH2023101869}. This approach enhances the extraction of the target speaker’s speech based on frequency characteristics while reducing dependency on language.

The input waveform $\mathbf{x}\in \mathbb{R}^L$ is transformed into a complex spectrogram $\mathbf{S} \in\mathbb{C}^{F \times T}$, where $F$ is the number of frequency bins and $T$ is the number of time frames in an utterance, respectively. $\mathbf{S}$ is equivalently represented as a real-valued feature map $\mathbf{U}_S \in \mathbb{R}^{2\times F \times T}$. Then $\mathbf{U}_S$ is transformed into $\mathbf{U}'_{S} \in \mathbb{R}^{D\times F \times T} $ via the 2D Convolution layer and Global Layer Normalization to create a high-dimensional representation. Each vector in D-dimensional vector space in $\mathbf{U}'_{S}$ encapsulates the speech features at a specific time and frequency bin.

The separation module employs the TF-GridNet block \cite{ZhongQiuWang2023TF_gridnet}, which comprises three primary components: the Intra-Frame Full-Band Module, the Sub-Band Temporal Module, and the Cross-Frame Self-Attention Module.
The Intra-Frame Full-Band Module utilizes a Bidirectional Long Short-Term Memory (BLSTM) network, treating the input spectrogram as a batch of $T$ signals, each of length $F$. This module is designed to capture features across different frequency bins within the same time frame. Next, the Sub-Band Temporal Module follows a similar architecture but swaps the roles of $F$ and $T$, enabling it to analyze the temporal evolution of features within each frequency band. Finally, the Cross-Frame Self-Attention Module incorporates a multi-head self-attention mechanism, aggregating information across multiple time frames. This module enhances the model’s ability to capture long-range dependencies in the temporal domain by effectively leveraging the attention mechanism.
\subsubsection{Speaker Fusion Module}
\begin{figure}[t]
  \centering
  \includegraphics[width=\linewidth]{images/whyv.pdf}
  \caption{Proposed architecture of WHYV Net: The input waveform is converted into a complex spectrogram via STFT. The spectrogram is processed using the TF-GridNet block as a separation module. The speaker embedding is mapped to GTF, and GTB serves as the input to the Speaker Fusion Module, which adjusts the spectrogram to match the target speaker. Finally, the spectrogram is reverted back into a waveform using iSTFT.}
  \label{fig:whyv}
  \vspace{-5 mm}
\end{figure}
The Speaker Fusion Module comes after the Separation Module to form a single \textit{WHYV block}, which is stacked $N$ times to progressively refine the representation of the target speaker. 
The Speaker Fusion Module transforms the output of the Separation Module to a deeper representation, incorporating additional speaker-specific information to enhance speech extraction.

In the WHYV design, a pretrained Voice Encoder extracts an embedding vector $\mathbf{e} \in \mathbb{R}^{D_e}$ (where $D_e$ is the embedding dimensions) from a speaker's reference audio. This embedding is then transformed into two frequency-domain speaker conditions: a Global Target Filter $\textbf{GTF} \in \mathbb{R}^{D\times F}$ and a Global Target Bias $\textbf{GTB} \in \mathbb{R}^{D\times F}$. These global conditions ensure consistent speaker conditions throughout the entire architecture. This approach avoids inconsistencies that could arise from directly injecting the embedding into each individual block, which may introduce variations in the learned condition.

The $k$-th Speaker Fusion Module takes the output of the $k$-th Separation Module, $\mathbf{I}_{k} \in \mathbb{R}^{D \times F \times T}$, as input, along with the global speaker conditions $\textbf{GTF} \in \mathbb{R}^{D \times F}$ and $\textbf{GTB} \in \mathbb{R}^{D \times F}$. To account for variability in how speaker information is abstracted within each block, each Speaker Fusion Module incorporates two learnable parameters: $\mathbf{\alpha} \in \mathbb{R}^{D \times F}$ and $\mathbf{\beta} \in \mathbb{R}^{D \times F}$. These parameters, sharing the same dimensions as the global conditions, allow adaptive conditioning within each block. The global conditions are then transformed into local conditions, $\mathbf{f}_k$ and $\mathbf{b}_k$ as follows:
\begin{equation}
\begin{aligned} \label{eq:gtf_gtb}
    \mathbf{f}_k &= \text{Sigmoid} \left( \mathbf{\alpha} \odot \textbf{GTF}\right),\\
    \mathbf{b}_k &= \text{Tanh} \left( \mathbf{\beta} \odot \textbf{GTB}\right), 
\end{aligned}    
\end{equation}
where $\odot$ denotes element-wise multiplication.

Next, the output of the $k$-th Speaker Fusion Module, $\mathbf{o}_k \in \mathbb{R}^{D \times F \times T}$, is defined as:
\begin{equation}
    \mathbf{o}_k = (\mathbf{f}_k \otimes \mathbf{1}_T) \odot \mathbf{I}_{k} + (\mathbf{b}_k \otimes \mathbf{1}_T) 
\end{equation}
where $\mathbf{1}_T$ is a 1D tensor of ones with size $T$, $\otimes$ represents the outer product, and $\odot$ denotes element-wise multiplication. 

Here, $\mathbf{f}_k$ acts as a speaker-dependent frequency gate. Through element-wise multiplication with $\mathbf{I}'_{k}$, it selectively amplifies or attenuates specific frequency components of the feature vector, effectively emphasizing or suppressing certain frequency bands characteristic of the target speaker. Meanwhile, $\mathbf{b}_k$ acts as a bias term, shifting the feature vector toward a representation that better captures the target speaker’s characteristics.


\section{Experiment}
\subsection{Training Configuration}
The WHYV Network is trained using the Libri2Mix dataset \cite{cosentino2020librimix} on a single NVIDIA A100 GPU with 80 GB of VRAM. Specifically, the \texttt{train-clean-100} subset is used, with training conducted under noisy conditions by incorporating noise from the WHAM! dataset \cite{Wichern2019WHAM}. Training is performed with a batch size of 16, using 4-second audio segments sampled at 8000 Hz. The AdamW optimizer \cite{loshchilov2018decoupled} is employed with a learning rate of $5 \times 10^{-4}$ and a weight decay of $1 \times 10^{-3}$. The loss function is based on the Scale-Invariant Signal-to-Distortion Ratio (SI-SDR) \cite{Jonathan2019SDR}, using the scale source variant as implemented in the TF-GridNet backbone \cite{Fengyuan24X_TF_GridNet}. We trained WHYV for 40 epochs. Additionally, the model can be trained on an NVIDIA A100 GPU with 40 GB of VRAM by halving the batch size to 8 while keeping all other parameters unchanged.

We utilize Resemblyzer\footnote{\url{https://github.com/resemble-ai/Resemblyzer}} as Voice Encoder, which is trained with a generalized end-to-end loss function for speaker verification \cite{Wan18Generalized}. Resemblyzer encodes a speech signal into a 256-dimensional vector $e \in \mathbb{R}^{256}$, representing the speaker's identity.
The Short-Time Fourier Transform (STFT) configuration and model parameters are shown in Table~\ref{tab:param}.

\subsection{Evaluation dataset}
To evaluate the performance of the WHYV network against other target speaker extraction (TSE) models, we conduct comparisons on the Libri2Mix test set. To demonstrate the language-agnostic generalization capability of the models, we prepare two non-English datasets for zero-shot evaluation. All models are trained exclusively on English data and evaluated on these datasets without further fine-tuning.

For the evaluation of non-English languages, we select Vietnamese and Mandarin due to the difference in accent and phonetic features. 
In addition, we don't have many suitable datasets available for the Vietnamese language, which makes it a low-resource language for the TSE problem.
We collect Vietnamese audio samples from publicly available sources to construct the evaluation set. The dataset consists of 20 speakers (10 male and 10 female) and approximately 13.4 hours of mixed audio. For Mandarin, we create the AISHELL3-2mix dataset from the AISHELL3 dataset \cite{AISHELL3_2020} to evaluate model performance. The AISHELL3-2mix dataset includes two versions: a clean version (AISHELL CLEAN) containing only two-speaker mixtures and a noisy version (AISHELL NOISY) augmented with background noise from the WHAM! dataset \cite{Wichern2019WHAM}. There are 214 different speakers in the test dataset of AISHELL3-2mix.

\begin{table}[t]
  \caption{STFT and Model Parameters}
  \label{tab:param}
  \centering
  \begin{tabular}{lcc}
    \toprule
    \textbf{Parameter Type} & \textbf{Parameter} & \textbf{Value} \\
    \midrule
    STFT Parameters & \texttt{n\_fft} & 128 \\
                    & \texttt{window\_size} & 128 \\
                    & \texttt{hop\_length} & 64 \\
    \midrule
    Model Parameters & $N$ (Number of blocks) & 5 \\
                     & $D$ (Feature dimensions) & 48 \\
                     & $E$ (Embedding dimensions) & 256 \\
                     & $H$ (BLSTM hidden size) & 192 \\
    \bottomrule
  \end{tabular}
  \vspace{-3mm}
\end{table}
\subsection{Result}
The WHYV model is evaluated on the Libri2Mix dataset and further tested in a zero-shot setting on an unseen language dataset. The detailed results are summarized in Table~\ref{tab:model_performance}. On the clean Libri2Mix dataset, WHYV achieves a Signal-to-Distortion Ratio (SDR) of 18.1 dB and a Perceptual Evaluation of Speech Quality (PESQ) score of 3.6. On the noisy version, it attains a SDR of 13.8 dB and a PESQ of 2.9.  
\begin{table}[t]
\centering
\caption{Performance of the WHYV model across diverse evaluation datasets, including clean and noisy conditions. Zero-shot evaluation is performed on AISHELL (Mandarin) and Vietnamese datasets, with results measured using SDR, SISDR, and PESQ metrics.}
\label{tab:model_performance}
\begin{tabular}{lccc}
\toprule
Dataset & \textbf{SDR} & \textbf{SISDR} & \textbf{PESQ} \\
\midrule
\textbf{LIBRI2MIX Clean} & 18.1 & 17.4 & 3.6 \\
\textbf{LIBRI2MIX Noisy} & 13.8 & 13.3 & 2.9 \\
\midrule
\textbf{VIETNAMESE Clean} & 14.9 & 14.6 & 3.2 \\
\midrule
\textbf{AISHELL Clean} & 13.3 & 13.4 & 3.2 \\
\textbf{AISHELL Noisy} & 10.1 & 10.2 & 2.6 \\
\bottomrule
\end{tabular}
\vspace{-2mm}
\end{table}

Table~\ref{tab:campare_others} compares the performance of WHYV with other recent methods on the Libri2Mix noisy benchmark for single-channel mixture audio processing. The results indicate that WHYV outperforms the majority of target speech extraction (TSE) models. However, some metrics were not reported in previous studies, limiting direct comparisons.

\begin{table}[b]
    \centering
    \caption{Comparison of the WHYV model with existing methods on the Libri2Mix noisy dataset, evaluated using SDR, SISDR, and SDRi metrics.}
    \label{tab:campare_others}
    \footnotesize 
    \setlength{\tabcolsep}{4pt} 
    \begin{tabular}{l S[table-format=2.2] S[table-format=2.4] S[table-format=2.3] S[table-format=2.3]}
        \toprule
        Model & {\# params (M)} & {SI-SDR} & {SDR} & {SDRi} \\
        \midrule
        WHYV (our) & 8.3 & 13.3 & 13.8 & 12.8 \\
        \midrule
        SpeakerBeam-SS \cite{Hiroshi24speakerbeam_ss} & 7.9 & \text{-} & 11.6 & \text{-} \\
        TSE Diffusion \cite{kamo23_interspeech} & \text{-} & 11.3 & \text{-} & \text{-} \\
        TSE with CL \cite{YunLiu24Target} & \text{-} & \text{-} & \text{-} & 13.5 \\
        \bottomrule
    \end{tabular}
    \vspace{-2mm}
\end{table}

\begin{table*}[t!]
  \caption{Comparison of Model Performance on the Libri2Mix Dataset and Zero-Shot Generalization to Untrained Languages (Vietnamese and Mandarin)}
  \label{tab:performance}
  \centering
  \footnotesize 
  \setlength{\tabcolsep}{2pt} 
  \begin{tabular}{l|c|ccc|ccc|ccc|ccc}
    \toprule
    \multirow{2}{*}{Method} & \multirow{2}{*}{Training Dataset} & \multicolumn{3}{c|}{LIBRI2MIX NOISY} & \multicolumn{3}{c|}{VIETNAMESE} & \multicolumn{3}{c|}{AISHELL CLEAN} & \multicolumn{3}{c}{AISHELL NOISY} \\
    \cmidrule(lr){3-5} \cmidrule(lr){6-8} \cmidrule(lr){9-11} \cmidrule(lr){12-14}
    & & SI-SDR↑ & SDR↑ & PESQ↑ & SI-SDR↑ & SDR↑ & PESQ↑ & SI-SDR↑ & SDR↑ & PESQ↑ & SI-SDR↑ & SDR↑ & PESQ↑ \\
    \midrule
    \textbf{WHYV (our)} & Libri2Mix & \textbf{13.3} & \textbf{13.8} & \textbf{2.9} & \textbf{14.6} & \textbf{14.9} & \textbf{3.2} & \textbf{13.4} & \textbf{13.3} & \textbf{3.2} & \textbf{10.2} & \textbf{10.1} & \textbf{2.6} \\
     & WHAMR! 
     & 8.9 & 8.8 & 2.2 & 7.7 & 6.7 & 2.3 & 10.3 & 10.1 & 2.7 & 7.4 & 7.1 & 2.1 \\
    \hline
    SpeakerBeam \cite{Zmolikova_Spkbeam_STSP19} & Libri2Mix & 11.2 & 11.5 & 2.4 & 11.1 & 11.3 & 2.4 & 7.8 & 7.1 & 2.1 & 6.5 & 5.5 & 1.8 \\
    X-TF-Gridnet \cite{Fengyuan24X_TF_GridNet} & WHAMR! & 7.2 & 7.9 & 2.2 & 5.8 & 5.1 & 2.0 & 5.2 & 6.7 & 2.2 & 4.1 & 4.0 & 1.8 \\
    BSRNN \cite{wang24fa_interspeech,Luo2023Music} & VoxCeleb1 & 2.9 & -1.3 & 1.5 & 4.6& 1.9 & 1.7 & 4.7 &1.5 & 1.7 & 2.8 & -1.4 & 0.6\\
    \bottomrule
  \end{tabular}
\end{table*}
In addition, Table~\ref{tab:performance} compares WHYV with other methods in the context of language transfer. Experiments are conducted using checkpoints provided by the authors of the competing methods, with reference audio lengths set according to the values reported in their respective papers. For WHYV, we develop two versions of the model: one trained on the Libri2Mix dataset and the other on the WHAMR! dataset. The results demonstrate that WHYV trained on Libri2Mix outperforms all competing approaches, underscoring its superior ability to generalize and effectively extract target speech across diverse languages. The version trained on WHAMR! maintains stable performance across varying conditions.

\begin{figure}[t]
  \centering
  \includegraphics[width=\linewidth]{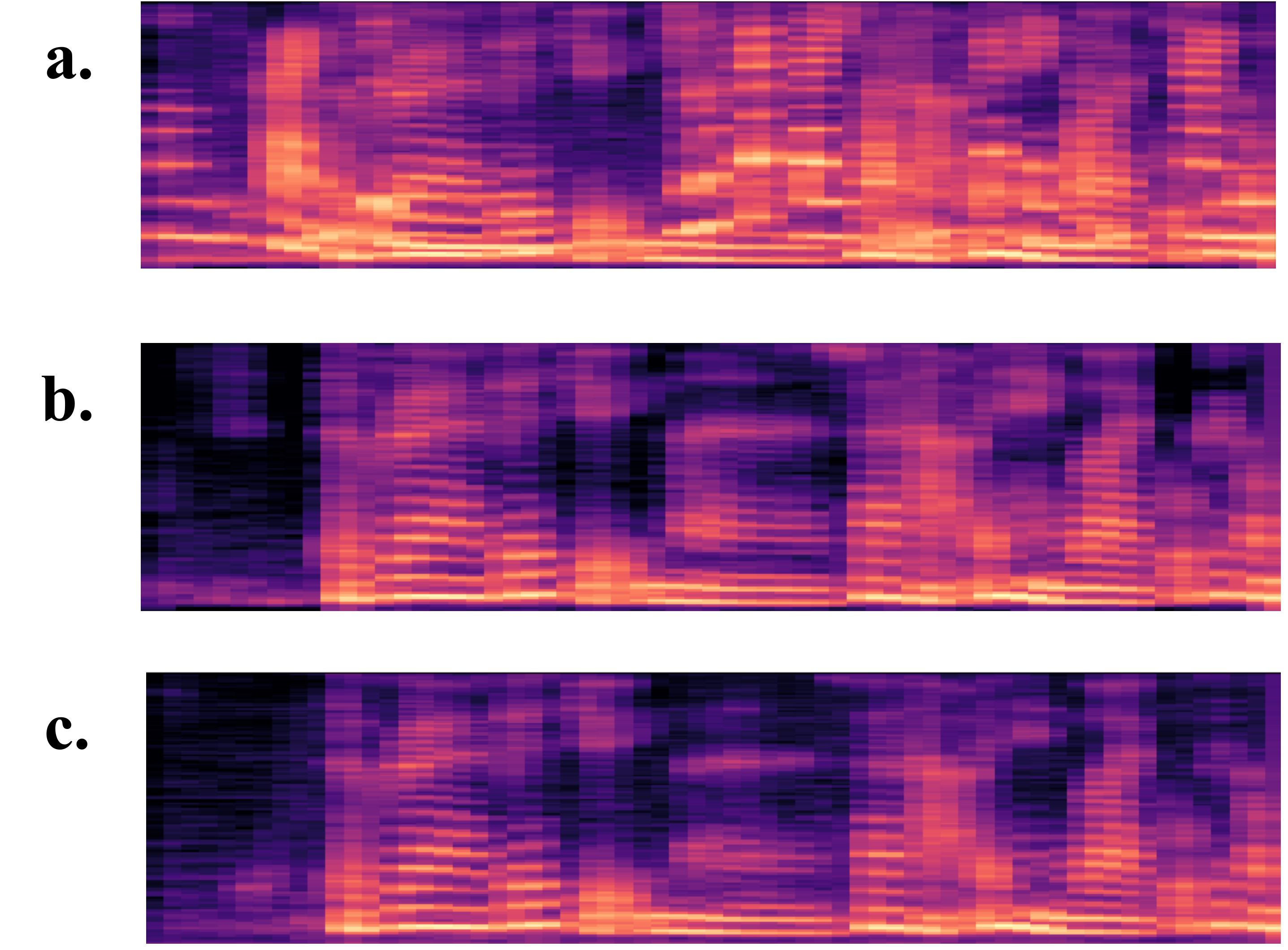}
  \caption{The spectrograms comparison between WHYV and SpeakerBeam across Libri2Mix, AISHELL, and Vietnamese datasets.  Note the erroneous speaker speech segments generated by SpeakerBeam when processing Mandarin (AISHELL), as highlighted.}
  \label{fig:spectrogram}
  \vspace{-5mm}
\end{figure}

Figure~\ref{fig:spectrogram} illustrates that SpeakerBeam, a competing method, struggles to differentiate between speakers when adapting to a new language. Specifically, when evaluated on Mandarin, it generates speech outputs that include segments from non-target speakers, as highlighted in the spectrogram. In contrast, WHYV exhibits robust adaptation capabilities. 

\subsection{Ablation Study}

Experiments are conducted to evaluate the contribution of the GTF \& GTF with learnable parameters in the Speaker Fusion Module. Alternative approaches have also been implemented, with results shown in Table~\ref{tab:gate_ablation}.

We conduct ablation studies to analyze the role of the Speaker Fusion module with Global Target Bias (GTB) and Global Target Filter (GTF).
In doing so, we replace the Speaker Fusion Module with other approaches: (1) a FiLM layer \cite{Perez2018Film} conditioned on speaker embeddings, (2) cross-attention, where queries are the target speaker's embeddings generated by the Voice Encoder, while key and value come from the Separation Module's output, (3) mutual attention \cite{hu2024smma_net} replacing standard attention. 

Results indicate that FiLM layers generalize poorly to unseen speakers despite enabling basic conditioning. In contrast, the proposed WHYV framework—leveraging GTF/GTB parameterization—ensures robust speaker consistency. Attention mechanisms underperform due to the compressed discriminative nature of pre-trained embeddings, which limits their ability to establish meaningful feature relationships. These findings emphasize the superiority of explicit conditioning via GTF/GTB over implicit attention-based fusion for speaker separation tasks.

Moreover, we conduct an experiment to investigate the role of learnable parameters in the Speaker Fusion Module. Since each block operates under different conditions due to varying levels of data abstraction, the technique used to adjust these conditions is crucial for maintaining consistency. As shown in Table~\ref{tab:linear}, we explore different methods to create the local conditions, including using the GTF/GTB with learnable parameters, as proposed in our approach, and compare it with alternative techniques. The first method applies a linear transformation, where global conditions are concatenated along the feature dimension. In the second approach, instead of using learnable parameters, we employ pointwise convolution to generate local conditions.

The results demonstrate that the learnable parameters when applied through pointwise multiplication, provide the most stable and consistent performance. In contrast, other techniques tend to create a black-box condition localized to each block, leading to inconsistencies in the conditioning. This highlights the importance of our proposed method in ensuring stable and coherent conditioning across different blocks.
\begin{table}[t]
    \centering
    \caption{Ablation study on the contribution of Speaker Fusion Module using GTB and GTF}
    \begin{tabular}{lc}
    \hline
    \textbf{Speaker Fusion Module} & \textbf{SI-SDR}  \\
    \hline
    \textbf{Ours} & 13.3  \\
    (1) FiLM \cite{Perez2018Film} & 10.2  \\
    (2) Attention \cite{vaswani2017attention} & 6.5  \\
    (3) Mutual attention \cite{hu2024smma_net} & 9.1 \\
    \hline
    \end{tabular}
    
    \label{tab:gate_ablation}
\end{table}

\begin{table}[t]
    \centering
    \caption{Ablation study on the learnable parameters using pointwise multiplication}
    \begin{tabular}{lc}
    \hline
    \textbf{Local Condition Transform Techniques} & \textbf{SI-SDR}  \\
    \hline
    \textbf{Ours} & 13.3  \\
    Linear Transformation & 10.5  \\
    Pointwise Convolution & 4.8 \\
    \hline
    \end{tabular}
    
    \label{tab:linear}
\end{table}

\section{Conclusion}
In this work, we propose the WHYV model for target speaker extraction using reference voice. The WHYV architecture introduces a new adaptive scheme that allows the model to utilize acoustic features in the separation process selectively. This approach demonstrates the ability to adapt to different languages, enabling the model to learn speaker-specific voice characteristics more effectively.
WHYV exhibits superior domain transfer capabilities compared to other models. 
To the best of our knowledge, it represents the state-of-the-art in cross-language domain transfer for target speaker extraction.

\newpage

\bibliographystyle{IEEEtran}
\bibliography{mybib,shortstring}

\end{document}